\documentclass[aps,pra,twocolumn,groupedaddress]{revtex4-1}

\usepackage{graphicx}
\usepackage{latexsym}
\usepackage{bm,array}
\usepackage{amsfonts}
\usepackage{amssymb}
\usepackage{amsmath}
\usepackage{upgreek}
\usepackage{xcolor}

\newcommand{\mtx}[2]{\left(\begin{array}{#1}#2\end{array}\right)}

\begin{document}


\title{Transition probabilities and transition rates in discrete phase space}


\author{William F.~Braasch Jr.}
\affiliation{Department of Physics and Astronomy, Dartmouth College, Hanover, New Hampshire 03755, USA}

\author{William K.~Wootters}
\affiliation{Department of Physics, Williams College, Williamstown, Massachusetts 01267, USA}



\begin{abstract}
The evolution of the discrete Wigner function is formally similar to a probabilistic process,
but the transition probabilities, like the discrete Wigner function itself, can be negative.  We investigate these transition probabilities, as well as the transition rates for a continuous process, aiming particularly to give 
simple criteria for deciding when a set of such quantities corresponds to a legitimate quantum process.  We also 
show how the transition rates for any Hamiltonian evolution can be worked out by expanding the Hamiltonian
as a linear combination of displacement operators in the discrete phase space.  
\end{abstract}


\maketitle


\section{Introduction}

The Wigner function is a real function on phase space representing the quantum state of a system of particles.  In Wigner's original paper, he points out that the equation of evolution of this function can be written in a form that makes the deterministic dynamics
look like a probabilistic process \cite{Wigner}.  Specifically, for each set of positions of the particles, and for each set of possible shifts in the particles' momenta, there is a certain probability
per unit time that the particles' momenta will undergo the specified shifts.  These transition rates, like the Wigner function itself, can be negative, so one cannot interpret the equation as literally representing a probabilistic process in the usual sense.  Elsewhere in the paper, though, Wigner observes that one will be able to obtain valid results by manipulating negative probabilities mathematically just as one would manipulate ordinary probabilities.  Several decades later, 
Feynman similarly argued that one should not automatically rule out the use of negative probabilities, again because the
end result can be perfectly sensible even if certain intermediate steps are difficult if not impossible to interpret \cite{Feynman}.  

The original Wigner function has now been extended, in a few different ways, to systems with a finite-dimensional Hilbert space.  One well-developed approach preserves the continuous nature of the phase space but gives it a shape and geometry appropriate to the system under study \cite{Stratonovich, Gracia, Varilly, Brif1, Brif2, Amiet, KlimovEspinoza, Tilma, Koczor1, Koczor2}.  For a single spin, with any Hilbert space dimension, the phase space is taken to be a two-dimensional spherical surface, matching the set of possible states of the analogous classical system.  A different approach---the one we follow in this paper---makes the phase space a discrete lattice with a size that depends on the 
dimension of the Hilbert space \cite{Buot, Hannay, Wootters, Galetti, Cohendet1, Cohendet2, Leonhardt1, Leonhardt2, Vourdas1, Luis, Hakioglu, Rivas, Gibbons, Vourdas2, Klimov, Chaturvedi, Gross1, Gross2, Chaturvedi2, Vourdas3}. In this approach, the function representing a quantum state assigns a real number to each lattice point and is called a discrete Wigner function. Discrete Wigner functions have found interesting applications in studies of entanglement characterization \cite{Franco}, quantum teleportation \cite{Koniorczyk, Paz1}, quantum algorithms \cite{Bianucci, Miquel1, Miquel2}, quantum computation \cite{Galvao, Cormick}, error-correcting codes \cite{Paz2} and quantum state tomography \cite{Paz3, Marchiolli, Wootters2}. As with the original Wigner function, the evolution 
of the discrete Wigner function can be expressed in the form of a probabilistic process, again with possibly negative
transition rates \cite{Klimov, Ruzzi}.
Alternatively, one can make the transition rates non-negative
by adding more structure to the discrete phase space (either a new binary variable \cite{Cohendet1, Cohendet2} or a phase \cite{Hashimoto}).  

In this paper we explore further the formulation of quantum evolution in discrete phase space, for two distinct ways
of characterizing this evolution.  (i) For a general
normalization-preserving quantum transformation, that is, for a trace-preserving completely positive map, we express the transformation
in terms of (possibly negative) transition probabilities in the discrete phase space.  (ii) For the special case of
Hamiltonian evolution---that is, for a closed system---we express the evolution in terms of transition probabilities 
per unit time (closely related work can be found in Refs.~\cite{Cohendet1, Cohendet2, Klimov, Ruzzi, Hashimoto}).  These transition rates come directly from a 
discrete version of the Moyal bracket \cite{Groenewold, Moyal}.  In both cases, our 
main goal is to formulate simple criteria for deciding whether a given set of transition probabilities or transition
rates corresponds to a legitimate quantum process.  That is, it turns out that one is not free to choose any properly normalized 
(but possibly negative) transition probabilities or transition rates, and we want to know what the constraints are.  
With a specification of these constraints,
the Wigner-function formulation of the evolution of finite-state systems becomes self-contained,
not requiring any reference to Hilbert space or to probability amplitudes.  The identification of the constraints
is the main new contribution of this paper.  

The paper is organized as follows.  In Section II we specify the particular form of the discrete Wigner function that we will be using.  In Section III we consider the transition probabilities in phase space for a general
normalization-preserving quantum transformation and ask what constraints there are on these probabilities.  We also
show in that section exactly how these constraints are strengthened when the evolution is unitary.
We do a similar analysis in Section IV for the transition {\em rates} in the case of continuous Hamiltonian evolution. 
These transition rates can be computed easily if one expands the Hamiltonian 
as a linear combination of displacement operators in the discrete phase space, as we explain in Section
V.  Finally we present our conclusions in Section VI.  In Appendix A, we work out the explicit form of the 
four-point structure function, which figures prominently in the definition of the transition probabilities.  
Appendices B and C prove technical results useful for characterizing, respectively, the allowed sets of 
transition probabilities and transition rates.

\section{Discrete Wigner function}

For our analysis we use the discrete Wigner function defined in Ref.~\cite{Wootters}.  That Wigner function is simplest 
when the dimension of the Hilbert space is a prime number, and for simplicity in the present paper, we restrict our attention to that case.  (When the state-space dimension is composite, the system is, in effect, treated as a composite object.)  

The discrete phase space can be pictured as an $N \times N$ array of points, where $N$ is the dimension of the system's Hilbert space.  We will use Greek letters to label the points of phase space, and for the point $\alpha = (\alpha_1, \alpha_2)$, we will picture $\alpha_1$ and $\alpha_2$ as the horizontal and vertical coordinates, respectively, where each $\alpha_i$ takes the values $0, 1, \ldots, N-1$.    Because $N$ is prime, these values, together
with the operations of addition and multiplication mod $N$, 
constitute a finite field.  With the coordinate labels understood as elements of this field, the phase space acquires the structure of a toroidal array.  In this phase space one can identify exactly $N(N+1)$ lines, that is, solutions of linear equations
in $\alpha_1$ and $\alpha_2$, and these lines can be sorted into $N+1$ sets, each set consisting of $N$ parallel lines.  We call a complete set of parallel lines a ``striation'' \cite{Gibbons}. 

For any density matrix $\hat{\rho}$, the corresponding Wigner function is defined as 
\begin{equation}  \label{W}
W_\alpha = \frac{1}{N}\, \hbox{Tr} (\hat{A}_\alpha \hat{\rho}),
\end{equation}
where the operators $\hat{A}_\alpha$ are given as follows. 
For $N=2$, 
\begin{equation}  \label{A2}
\hat{A}_\alpha = \frac{1}{2}[\hat{I} + (-1)^{\alpha_1} \hat{Z} + (-1)^{\alpha_2} \hat{X} + (-1)^{\alpha_1 + \alpha_2} \hat{Y},
\end{equation}
where $\hat{I}$ is the $2 \times 2$ identity matrix and $\hat{X}$, $\hat{Y}$, and $\hat{Z}$ are the Pauli matrices (to be generalized below).  For prime $N$ greater 
than 2, we write $\hat{A}_\alpha$ in terms of the phase-space displacement operators $\hat{D}_\beta$, defined by \cite{Schwinger,Vourdas2}
\begin{equation} \label{displacements}
\hat{D}_\beta = \omega^{\beta_1 \beta_2/2} \hat{X}^{\beta_1} \hat{Z}^{\beta_2}.
\end{equation}
Here $\omega = e^{2 \pi i/N}$ and the arithmetic in its exponent is understood to be mod $N$.  (So $\omega^{1/2} = \omega^{(N+1)/2}$.)  The basic displacement operators 
$\hat{X}$ and $\hat{Z}$---generalized Pauli matrices---are defined in terms of a standard orthonormal basis 
$\{|q\rangle \}$ as \cite{Weyl}
\begin{equation} \label{XZdef}
\begin{split}
&\hat{X} |q\rangle = |q+1 \; \hbox{(mod $N$)}\rangle \\
&\hat{Z} |q\rangle = \omega^q |q\rangle.
\end{split}
\end{equation}
Now we define the operators $\hat{A}_\alpha$ by
\begin{equation}  \label{DAs}
\hat{A}_\alpha = \frac{1}{N} \sum_\beta \hat{D}_\beta \omega^{\langle \alpha, \beta \rangle},
\end{equation}
where $\langle \cdot, \cdot \rangle$ denotes the symplectic product
$\langle \alpha, \beta \rangle = \alpha_2 \beta_1 - \alpha_1 \beta_2$.
In terms of its matrix components (in the standard basis), we can write $\hat{A}_\alpha$ as
\begin{equation}  \label{Aodd}
(\hat{A}_\alpha)_{kl} = \delta_{2\alpha_1, k+l}\, \omega^{\alpha_2 (k-l)},
\end{equation}
where the matrix indices $k$ and $l$ take the values $0, 1, \ldots, N-1$, and the arithmetic
in the subscript of the Kronecker delta is mod $N$.  

The Hermitian operators $\hat{A}_\alpha$, which we call phase-point operators (they are also called Fano operators \cite{Cohendet1}), have a number of special properties:

\begin{quote}
(i) $ \hbox{Tr}\, \hat{A}_\alpha = 1.$
\end{quote}
\begin{quote}
(ii) $ \hbox{Tr}  (\hat{A}_\alpha \hat{A}_\beta) = N \delta_{\alpha\beta}.$
\end{quote}
\begin{quote}
(iii) For any striation consisting of lines $\lambda$, the operators $\hat{Q}_\lambda = (1/N)\sum_{\alpha \in \lambda} \hat{A}_\alpha$ are projection operators onto the elements of an orthonormal basis of the Hilbert space.  Moreover, the
bases corresponding to different striations are mutually unbiased; that is, if the lines $\lambda_1$ and $\lambda_2$ are not parallel, then $\hbox{Tr}  (\hat{Q}_{\lambda_1}\hat{Q}_{\lambda_2}) = 1/N$.
\end{quote}
\begin{quote}
(iv) As follows immediately from (iii), $(1/N)\sum_\alpha \hat{A}_\alpha = \hat{I}$, where $\hat{I}$ is the identity.    
\end{quote}
For $N=2$, statements (i) and (ii) can be proven using compositional properties of the Pauli operators. The same statements can be verified directly for odd prime $N$ by replacing each $\hat{A}$ with its definition (\ref{DAs}) and 
using the fact that
\begin{equation}  \label{DtimesD}
\hat{D}_\alpha \hat{D}_\beta = \omega^{\langle\alpha,\beta\rangle/2} \hat{D}_{\alpha+\beta}.
\end{equation}
This multiplication rule for the displacement operators follows from Eqs.~(\ref{displacements}) and (\ref{XZdef}) via the commutation relation ${\hat{X}^n \hat{Z}^m = \omega^{-mn} \hat{Z}^m \hat{X}^n}$ \cite{Schwinger, Vourdas2}. Finally, one can obtain statement (iii) (and thus also statement (iv)) from Eqs. (\ref{A2}) and (\ref{Aodd}) by explicitly summing over the lines of the discrete phase space to find the operators $Q_\lambda$ \cite{Wootters}.

The second of the above statements expresses the fact that the $\hat{A}$'s constitute an orthogonal basis 
for the space of $N \times N$ matrices, so that we can write any such matrix as a linear combination
of the $\hat{A}$'s.  In particular, we can invert Eq.~(\ref{W}):
\begin{equation}
\hat{\rho} = \sum_\alpha W_\alpha \hat{A}_\alpha.
\end{equation}
That is, the values of the Wigner function are simply the coefficients in the expansion of $\hat{\rho}$ in the 
phase-point operators.
Meanwhile the first and third of the above statements imply the following properties of the Wigner function.  
\begin{quote}
(a)  $\sum_\alpha W_\alpha = 1$.
\end{quote}
\begin{quote}
(b)  The sums of $W_\alpha$ over the lines of a striation are the probabilities of the outcomes of the orthogonal
measurement associated with that striation.  
\end{quote}
These properties, which are analogous to properties of the continuous Wigner function, provide a sense in which the 
discrete Wigner function acts like a probability distribution: the Wigner function is normalized like a probability distribution, and the marginal distribution over each direction in phase space is an actual, non-negative probability distribution, corresponding to a complete orthogonal measurement.  For example, for the spin of a spin-1/2 particle,
with $N=2$, the three marginals (over the horizontal, diagonal, and vertical lines) can be interpreted as the probability distributions for spin measurements along the $x$, $y$, and $z$ axes \cite{Wootters, Feynman}.  However, like the continuous Wigner function, $W_\alpha$ can take negative
values.  Indeed, for the case $N=2$, the Wigner function we are using is essentially the same as a function
Feynman defined in one of his examples of negative probabilities \cite{Feynman}.

For a particle moving in one continuous dimension, we usually interpret the axis variables of phase space as position and momentum.  In our discrete case, the interpretation of the axis variables will depend on the particular system under study; e.g., the horizontal axis may be associated with values of the $z$-component of spin.  The example we give in Section V---a particle confined to a discrete ring of possible locations---is probably the discrete system most closely analogous to the continuous case.  There we interpret the horizontal axis variable as position and the vertical axis variable as the discrete wavenumber, which is analogous to momentum.  In general, though, the results we present in this paper are independent of the interpretation of the axes.  

Not every normalized real function on phase space corresponds to an actual quantum state.  
One way of identifying the legitimate functions $W_\alpha$ is simply to say they are the ones for which
$\sum_\alpha W_\alpha \hat{A}_\alpha$ is a positive semidefinite matrix.  Another way is to focus first on pure states.
Recall the property $\hat{\rho} = \hat{\rho}^2$ of pure state density matrices.
Recasting this as a discrete phase space expression, one finds that the pure states are represented by normalized functions $W_\alpha$ satisfying
\begin{equation}  \label{pureW}
W_\alpha = \sum_{\beta\gamma} \Gamma_{\alpha\beta\gamma} W_\beta W_\gamma,
\end{equation}
where $\Gamma_{\alpha\beta\gamma}$ is the three-point structure function
\begin{equation}  \label{Gamma}
\Gamma_{\alpha\beta\gamma} = \frac{1}{N} \hbox{Tr}(\hat{A}_\alpha \hat{A}_\beta \hat{A}_\gamma).
\end{equation}
Mixed states can then be identified as the convex combinations of pure states \cite{Wootters}.  

Later we will use the following symmetries of $\Gamma_{\alpha\beta\gamma}$:
\begin{equation} \label{Gamma-props}
\begin{split}
&\hbox{(a)} \;\; \Gamma_{\alpha+\delta, \beta+\delta, \gamma+\delta} = \Gamma_{\alpha \beta\gamma}.  \\
&\hbox{(b)} \;\; \Gamma_{\alpha\beta\gamma} = \Gamma_{\gamma\alpha\beta} = \Gamma^*_{\alpha\gamma\beta},
\end{split}
\end{equation}
where $\delta$ is any ordered pair $(\delta_1, \delta_2)$ and the asterisk indicates complex conjugation. For odd $N$, property (a) in Eq.~(\ref{Gamma-props}) can be proven by expressing the phase-point operators in terms of the unitary displacement operators using Eq.~(\ref{DAs}) and then observing, via the multiplication rule (\ref{DtimesD}), that $\hat{A}_{\alpha+\delta} = \hat{D}_\delta \hat{A}_\alpha \hat{D}_\delta^\dag$.  This last equation holds also for $N=2$ with the Pauli operators playing the role of the $\hat{D}$'s.  Property (b) in Eq.~(\ref{Gamma-props}) follows from the cyclic property of the trace and the fact that for any square matrix $\hat{M}$, $(\hbox{Tr}\, {\hat{M}})^* = \hbox{Tr} (\hat{M}^\dag)$.

Finally, we note here two further properties of the $\hat{A}$ operators that we will find useful.
For any $N \times N$ matrix $\hat{M}$,
\begin{equation} \label{properties}
\begin{split}
&\hbox{(a)} \;\; \sum_\alpha \hat{A}_\alpha \hbox{Tr}(\hat{M}\hat{A}_\alpha) = N\hat{M} \\
&\hbox{(b)} \;\; \sum_\alpha \hat{A}_\alpha \hat{M} \hat{A}_\alpha = N(\hbox{Tr}\hat{M})\hat{I}.
\end{split}
\end{equation}
Both of these equations follow directly from the orthogonality and normalization of the phase-point operators. 
Eq.~(\ref{properties}b) can be proved
as follows.  First consider the simple orthonormal
basis $\hat{E}_{\alpha} = |j\rangle\langle k|$ for the space of $N \times N$ matrices, where $\alpha$ 
stands for the pair $(j,k)$.  One can show directly
that $\sum_\alpha \hat{E}_\alpha \hat{M} \hat{E}_\alpha^\dag = (\hbox{Tr}\hat{M})\hat{I}$. 
Any other orthonormal basis $\hat{F}_\alpha$ can be written
as $\hat{F}_\alpha = \sum_\beta U_{\alpha\beta} \hat{E}_\beta$, where
$U$ is an $N^2 \times N^2$ unitary matrix.  It follows that
$\hat{F}_\alpha$ satisfies the same sum rule.  We get Eq.~(\ref{properties}b) by taking into account the different
normalization of the $\hat{A}$'s.

\section{Trace-preserving quantum operations}

Consider a quantum system ${\mathcal S}$ with Hilbert-space dimension $N$, possibly interacting with an environment.  As long as there is no initial correlation between the system and the environment, the most general 
transformation of ${\mathcal S}$ is represented by a completely positive map.  If this map takes an initial density matrix $\hat{\rho}$
of ${\mathcal S}$ to a final, normalized density matrix $\hat{\rho}'$ of the same system, it can be expressed in the form
\begin{equation} \label{transformation}
\hat{\rho}' = \sum_j \hat{B}_j \hat{\rho} \hat{B}^\dag_j,
\end{equation}
where the $N \times N$ Kraus matrices $\hat{B}_j$ satisfy the condition
\begin{equation} \label{Kraus}
\sum_j \hat{B}_j^\dag \hat{B}_j = \hat{I}.  
\end{equation}

It is a straightforward matter to re-express Eq.~(\ref{transformation}) as a transformation of the discrete Wigner function.  For each $\hat{B}_j$, let us define the corresponding phase-space function $B^{(j)}_{\alpha}$ by
\begin{equation}
B^{(j)}_{\alpha} = \frac{1}{N} \hbox{Tr}\, (\hat{A}_\alpha \hat{B}_j),
\end{equation}
so that 
$\hat{B}_j = \sum_\alpha B^{(j)}_{\alpha} \hat{A}_\alpha$.  
The condition (\ref{Kraus}) then becomes
\begin{equation}  \label{Bcondition}
\sum_{\alpha\beta} {\mathcal B}_{\alpha\beta}  \hat{A}_\beta \hat{A}_\alpha = \hat{I},
\end{equation}
where 
\begin{equation}  \label{Bdef}
{\mathcal B}_{\alpha\beta} = \sum_j {B}^{(j)}_{\alpha} {B}^{(j)*}_{\beta},
\end{equation}
Expanding $\hat{\rho}$, $\hat{\rho}'$ and the $\hat{B}$'s in Eq.~(\ref{transformation}),
we obtain
\begin{equation}
W'_\alpha = \frac{1}{N} \sum_{\beta\gamma\delta} \hbox{Tr}(\hat{A}_\alpha \hat{A}_\beta \hat{A}_\gamma \hat{A}_\delta) {\mathcal B}_{\beta\delta} W_\gamma.
\end{equation}
Thus by defining $P_{\alpha\gamma}$ to be
\begin{equation} \label{Pdef}
P_{\alpha\gamma} = \frac{1}{N} \sum_{\beta\delta} \hbox{Tr}(\hat{A}_\alpha \hat{A}_\beta \hat{A}_\gamma \hat{A}_\delta) {\mathcal B}_{\beta\delta},
\end{equation}
we can express the evolution as
\begin{equation} \label{probprocess}
W'_\alpha = \sum_\gamma P_{\alpha\gamma} W_{\gamma}.
\end{equation}
If we interpret $W_\gamma$ as the probability of finding the system at the phase-space point $\gamma$, then 
$P_{\alpha\gamma}$ plays the role of the probability that
a system at the point $\gamma$ will make a transition to $\alpha$ (but $P_{\alpha\gamma}$ can be negative).  In what follows,
we will save a bit of space by defining the four-point structure function
\begin{equation}
\Xi_{\alpha\beta\gamma\delta} = \frac{1}{N} \hbox{Tr}(\hat{A}_\alpha \hat{A}_\beta \hat{A}_\gamma \hat{A}_\delta).
\end{equation}
Then Eq.~(\ref{Pdef}) becomes
\begin{equation} \label{PXi}
P_{\alpha\gamma} = \sum_{\beta\delta} \Xi_{\alpha\beta\gamma\delta} {\mathcal B}_{\beta\delta}.
\end{equation}
We note for future reference that in addition to being invariant under cyclic permutations of its indices,
$\Xi_{\alpha\beta\gamma\delta}$ also has the following symmetry: 
\begin{equation}  \label{Xisymmetry}
\Xi_{\gamma\beta\alpha\delta} = 
\Xi_{\alpha\beta\gamma\delta}^*
\end{equation}
since switching $\alpha$ and $\gamma$ effectively reverses the order
of the $A$ operators inside the trace (and we again use the fact that $(\hbox{Tr}\, {\hat{M}})^* = \hbox{Tr} (\hat{M}^\dag)$). 

We can extend the definition of $P_{\alpha\gamma}$ to 
the case of linear transformations ${\mathcal E}$ that are not necessarily
completely positive.  Let Eq.~(\ref{probprocess}) serve as
the definition of $P_{\alpha\gamma}$ for such a transformation.
Then from $\hat{\rho}' = {\mathcal E}(\hat{\rho})$ and Eq.~(\ref{probprocess}), we have
\begin{equation}
{\mathcal E}\Big(\sum_\beta W_\beta \hat{A}_\beta\Big) = \sum_\sigma W'_\sigma \hat{A}_\sigma
= \sum_{\sigma\tau} P_{\sigma\tau} W_\tau \hat{A}_\sigma.
\end{equation}
Now inserting for the initial Wigner function the illegal state $W_\beta = \delta_{\beta\gamma}$, we get
\begin{equation}  \label{EfromP}
{\mathcal E}(\hat{A}_\gamma) = \sum_\sigma P_{\sigma\gamma} \hat{A}_\sigma \; ,
\end{equation}
from which it follows that
\begin{equation}  \label{PfromE}
P_{\alpha\gamma} = \frac{1}{N} \hbox{Tr} [\hat{A}_\alpha {\mathcal E}(\hat{A}_\gamma)].
\end{equation}
(It is perfectly fine to use the illegal state $W_\beta = \delta_{\beta\gamma}$ in this derivation.
The map ${\mathcal E}$ is defined for all operators acting on the $N$-dimensional
Hilbert space, not just density operators.  This illegal Wigner function corresponds to the operator $\hat{A}_\gamma$.) 
The specific formula for $P_{\alpha\gamma}$ given in Eq.~(\ref{PXi}) follows from Eq.~(\ref{PfromE}) when 
${\mathcal E}$ can be expressed in the form (\ref{transformation}).

Now, if we were allowed to choose $P_{\alpha\gamma}$ arbitrarily, even if we were to insist on the 
standard normalization condition $\sum_\alpha P_{\alpha\gamma} = 1$, we would easily be able to 
create an illegal quantum state from a legal one.  For example, all points in phase space could be mapped
with probability 1 to a specific point.  Then according to property (b) of the Wigner function in Section II, the final state $W'$ would produce a deterministic outcome for each
of the $N+1$ mutually unbiased measurements associated with the striations, which is impossible.  So we now
ask this question: given a proposed set of transition probabilities $P_{\alpha\gamma}$, how does one know whether
it corresponds to a valid quantum transformation?  

We begin by inverting Eq.~(\ref{Pdef}) so as to express ${\mathcal B}_{\beta\delta}$ in terms of the $P$'s.  
The details of this inversion are found in Appendix B, with the result
\begin{equation}  \label{inverse}
{\mathcal B}_{\beta\delta} = \frac{1}{N^2} \sum_{\alpha\gamma} \Xi_{\beta\alpha\delta\gamma} P_{\alpha\gamma}.
\end{equation}
Comparing this equation to Eq.~(\ref{PXi}), we see that  
$P$'s and the ${\mathcal B}$'s are related to each other
in a symmetric way. One consequence of Eq.~(\ref{inverse}) is that the values ${\mathcal B}_{\beta\delta}$ are uniquely determined by the quantum transformation: according to Eq.~(\ref{probprocess}), specifying the transformation is equivalent to specifying the transition probabilities, and these probabilities in turn 
determine the ${\mathcal B}$'s through 
Eq.~(\ref{inverse}).  In this respect 
the ${\mathcal B}$'s differ from the set of operators $\hat{B}_j$, for which
one can choose among many different sets that all represent the same transformation. 

In the preceding paragraph, we began by assuming implicitly that the $P$ values we were given could be expressed in the
form (\ref{Pdef}).  But how do we know that for a given set of $P$ values, there exist a set of complex numbers 
${\mathcal B}_{\beta\delta}$ such that the $P$'s can be expressed in that form?  (In asking this question we are not yet insisting that the 
${\mathcal B}$'s arise from a legitimate set of $\hat{B}_j$ operators.)  In fact
this is not a problem.  For any numbers $P_{\alpha\gamma}$, if we insert the ${\mathcal B}$'s of Eq.~(\ref{inverse}) back into Eq.~(\ref{Pdef}), we find that we arrive again at the values of $P_{\alpha\gamma}$ that we started with. 
This is because 
\begin{equation}
\frac{1}{N^2} \sum_{\beta\delta} \Xi_{\alpha\beta\gamma\delta} \Xi_{\beta\sigma\delta\tau} = \delta_{\alpha\sigma}\delta_{\gamma\tau}\; ,
\end{equation}
as can be shown directly using the properties of the $A$'s given in Eq.~(\ref{properties}).   Thus any set of $P$'s
is consistent with Eq.~(\ref{Pdef}) if we allow the ${\mathcal B}$'s to be entirely unconstrained.  
 
Our first constraint on the $P$'s comes from Eq.~(\ref{Bcondition}), which places a condition
on ${\mathcal B}$.  Let us use Eq.~(\ref{inverse}) to express this condition in terms of the transition probabilities.  
Using Eq.~(\ref{properties}), we obtain
\begin{equation}
\begin{split}
\hat{I} &= \sum_{\mu\nu}{\mathcal B}_{\mu\nu} \hat{A}_\nu \hat{A}_\mu = \frac{1}{N^3}\sum_{\mu\nu\alpha\gamma}
P_{\alpha\gamma} \hbox{Tr}(\hat{A}_\mu \hat{A}_\alpha \hat{A}_\nu \hat{A}_\gamma) \hat{A}_\nu \hat{A}_\mu \\
&=\frac{1}{N^2}\sum_{\nu\alpha\gamma} P_{\alpha\gamma} \hat{A}_\nu \hat{A}_\alpha \hat{A}_\nu \hat{A}_\gamma \hspace{1cm} \hbox{[by Eq.~(\ref{properties}a)]} \\
&=\frac{1}{N}\sum_{\alpha\gamma} P_{\alpha\gamma} (\hbox{Tr}\, \hat{A}_\alpha ) \hat{A}_\gamma \hspace{1.3cm}
\hbox{[by Eq.~(\ref{properties}b)]} \\
&= \frac{1}{N} \sum_\gamma \left( \sum_\alpha P_{\alpha\gamma} \right) \hat{A}_\gamma.
\end{split}
\end{equation}
This condition will be satisfied as long as the $P$'s are normalized in the sense that $\sum_\alpha P_{\alpha\gamma} = 1$ for every $\gamma$.  Moreover, the equation {\em implies} this normalization condition, as can be seen by 
multiplying both sides by $\hat{A}_\tau$ and taking the trace.  Thus the condition (\ref{Bcondition}) is equivalent to the
natural normalization condition on the $P$'s.  

We get a more restrictive condition on the $P$'s from the form of the definition
of ${\mathcal B}_{\beta\delta}$.  Regarded as a matrix with $\beta$ and $\delta$ as the matrix indices,
we can see from Eq.~(\ref{Bdef}) that
${\mathcal B}_{\beta\delta}$ must be positive semidefinite: any matrix that can be written in this form is 
positive semidefinite, and any positive semidefinite matrix can be written in this form.   
Thus we arrive at our criteria for determining whether a given set of transition probabilities $P_{\alpha\gamma}$
represents a legitimate quantum process:
\begin{equation} \label{Pcondition}
\begin{split}
&\hbox{(a)} \;\; \sum_\alpha P_{\alpha\gamma} = 1 \hspace{2mm} \hbox{for every $\gamma$} \\
&\hbox{(b)} \;\; \sum_{\alpha\gamma} \Xi_{\beta\alpha\delta\gamma} P_{\alpha\gamma}\hspace{2mm} \hbox{is positive semidefinite},
\end{split}
\end{equation}
where the mathematical expression in Eq.~(\ref{Pcondition}b) is understood to be a matrix with indices $\beta$ and $\delta$.

As an example, consider the set of transition probabilities for a single qubit defined as follows:
\begin{equation}  \label{example}
P_{\alpha\gamma} = \frac{1}{2} - \delta_{\alpha+\gamma, \zeta} \; ,
\end{equation}
where $\zeta$ is the ordered pair $(1,1)$.  That is, $P$ has the value $1/2$ unless the transition is to the
opposite corner of the $2 \times 2$ phase space, in which case $P$ has the value $-1/2$.  These values of $P$ are 
properly normalized.  To check whether they represent an actual quantum transformation, we use
Eq.~(\ref{inverse}) and Eq.~(\ref{A2}) to find ${\mathcal B}_{\beta\delta}$.  The result is 
\begin{equation}
{\mathcal B} = \frac{1}{4} \mtx{cccc}{1 & 1 & 1 & -1 \\ 1 & 1 & -1 & 1 \\ 1 & -1 & 1 & 1 \\ -1 & 1 & 1 & 1},
\end{equation}
where the vertical and horizontal indices are interpreted as $\beta$ and $\delta$, respectively, each index taking the values 
$(0,0), (0,1), (1,0), (1,1)$ in that order.  This matrix has the eigenvalues $(1/2, 1/2, 1/2, -1/2)$ and is therefore not positive semidefinite.  So the transition probabilities defined in Eq.~(\ref{example}) do not correspond to a possible transformation on 
a qubit.  In fact, one can show from Eq.~(\ref{EfromP}) that they correspond to the transpose operation, which is the prototypical example
of a positive but not completely positive map.  

The condition (\ref{Pcondition}b) requires determining whether a certain $N^2 \times N^2$ matrix---the matrix ${\mathcal B}_{\beta\delta}$ given by Eq.~(\ref{inverse})---is positive 
semidefinite.  In this respect it is similar to a more standard test for complete positivity, namely, to see whether the
Choi operator, another $N^2 \times N^2$ matrix, is positive semidefinite \cite{Jamiolkowski, Choi, Wilde}.  In fact,
it turns out that
${\mathcal B}_{\beta\delta}$ is simply the Choi operator
written in a specific basis, as we now show.

From Eqs.~(\ref{PfromE}) and (\ref{inverse}), we have
\begin{equation}  \label{BwithE}
\begin{split}
    {\mathcal B}_{\beta\delta} &= \frac{1}{N^4}\sum_{\alpha\gamma}
    \hbox{Tr}(\hat{A}_\beta \hat{A}_\alpha \hat{A}_\delta \hat{A}_\gamma) \hbox{Tr}\left[\hat{A}_\alpha {\mathcal E}(\hat{A}_\gamma)\right]\\
    &=\frac{1}{N^3}\sum_\gamma \hbox{Tr}\left[\hat{A}_\delta \hat{A}_\gamma \hat{A}_\beta {\mathcal E}(\hat{A}_\gamma)\right].
\end{split}
\end{equation}
The Choi operator is
\begin{equation}  \label{Choidef}
    \hat{C} = \frac{1}{N}\sum_{jk} |j\rangle\langle k| \otimes {\mathcal E}(|j\rangle\langle k|).
\end{equation}
Let us define the orthonormal basis $|\Psi_\alpha \rangle$ by
\begin{equation}  \label{Psidef}
    |\Psi_\beta\rangle = (\hat{I} \otimes \hat{A}_\beta)|\Phi\rangle,
\end{equation}
where $|\Phi\rangle$ is the maximally entangled state
\begin{equation}
    |\Phi\rangle = \frac{1}{\sqrt{N}} \sum_m |m\rangle \otimes |m\rangle.
\end{equation}
Then we claim that
\begin{equation} \label{BCclaim}
    {\mathcal B}_{\beta\delta} = \langle \Psi_\beta|\hat{C}|\Psi_\delta\rangle.
\end{equation}
Indeed, by plugging the definitions (\ref{Choidef}) and (\ref{Psidef})
into the right-hand side of Eq.~(\ref{BCclaim}), we find that
\begin{equation} \label{firstB}
\begin{split}
   \langle \Psi_\beta|\hat{C}|\Psi_\delta\rangle &= 
   \frac{1}{N^2} \sum_{jk} \hbox{Tr}\left[ |k\rangle\langle j| \hat{A}_\beta {\mathcal E}(|j\rangle\langle k|) \hat{A}_\delta \right]\\
   &=\frac{1}{N^2} \sum_\alpha \hbox{Tr} \left[ \hat{E}_\alpha^\dag
   \hat{A}_\beta {\mathcal E}(\hat{E}_\alpha) \hat{A}_\delta \right],
\end{split}
\end{equation}
where we are defining $\hat{E}_\alpha$ to be $|j\rangle\langle k|$, with $\alpha = (j,k)$.  We know that we can write the orthonormal matrix basis
$\{\hat{E}_\alpha\}$ in terms of the alternative orthonormal matrix basis $\{\hat{A}_\gamma/\sqrt{N}\}$ as
\begin{equation}
    \hat{E}_\alpha = \sum_\gamma U_{\alpha\gamma}(\hat{A}_\gamma/\sqrt{N}),
\end{equation}
where $U$ is an $N^2 \times N^2$ unitary matrix.  (In fact, one can check that $U_{\alpha\gamma} = (1/\sqrt{N})\delta_{2\gamma_1,j+k}
\omega^{-\gamma_2(j-k)}$.)  It follows that we can replace
the basis $\{\hat{E}_\alpha\}$ in Eq.~(\ref{firstB}) with $\{\hat{A}_\gamma/\sqrt{N}\}$.  This gives us  
\begin{equation}
 \langle \Psi_\beta|\hat{C}|\Psi_\delta\rangle = \frac{1}{N^3}\sum_\gamma \hbox{Tr}\left[\hat{A}_\gamma \hat{A}_\beta {\mathcal E}(\hat{A}_\gamma)\hat{A}_\delta \right], 
\end{equation}
which agrees with Eq.~(\ref{BwithE}).  Thus ${\mathcal B}$ is 
the Choi operator written in the basis $|\Psi_\beta\rangle$.


The above analysis becomes simpler in the case of unitary evolution.  In that case, we can get an expression for
the transition probabilities directly from Eq.~(\ref{PfromE}):
\begin{equation} \label{unitaryP}
P_{\alpha\rho} = \frac{1}{N} \hbox{Tr}(\hat{A}_\alpha \hat{U} \hat{A}_\rho \hat{U}^\dag),
\end{equation}
where $U$ is the unitary evolution operator.  From Eq.~(\ref{unitaryP}) and Eq.~(\ref{properties}a), we 
can see that this $P_{\alpha\rho}$,
regarded as a matrix with indices $\alpha$ and $\rho$,
is an {\em orthogonal} matrix: $PP^T = I$.  We also note that in this case ${\mathcal B}_{\beta\delta}$ has the 
simple form
\begin{equation}  \label{unitaryB}
{\mathcal B}_{\beta\delta} = B_\beta {B}^*_\delta  \hspace{1cm} B_\beta = \frac{1}{N} \hbox{Tr}(\hat{U} \hat{A}_\beta),
\end{equation}
from which it follows that $\sum_\beta |B_\beta|^2 = 1$.  

In Ref.~\cite{Wootters} it was shown that, among all real functions of two phase-space points, those $P$'s that 
correspond to unitary transformations are completely characterized by the following 
two properties:
\begin{equation}  \label{oldPcondition}
\begin{split}
&\hbox{(a)} \;\; \sum_\alpha P_{\alpha\rho} = 1. \\
&\hbox{(b)} \;\; \sum_{\rho\sigma\tau} P_{\alpha\rho} P_{\beta\sigma} P_{\gamma\tau} \Gamma_{\rho\sigma\tau} = \Gamma_{\alpha\beta\gamma},
\end{split}
\end{equation}
where $\Gamma_{\alpha\beta\gamma}$ is the three-point structure function we defined in Eq.~(\ref{Gamma}).
That is, in addition to the standard normalization condition, the $P$'s must leave $\Gamma$ unchanged.  

In the spirit of Eq.~(\ref{Pcondition}), we can replace Eq.~(\ref{oldPcondition}b) with an alternative condition, so that the conditions for a unitary transformation become
\begin{equation}  \label{newPcondition}
\begin{split}
&\hbox{(a)} \;\; \sum_\alpha P_{\alpha\gamma} = 1. \\
&\hbox{(b)} \;\; {\mathcal B}_{\beta\delta} = \frac{1}{N^2} \sum_{\alpha\gamma} \Xi_{\beta\alpha\delta\gamma} P_{\alpha\gamma} \;\; \hbox{has rank 1},
\end{split}
\end{equation}
where again $\beta$ and $\delta$ are understood to be matrix indices.  
It is clear from Eq.~(\ref{unitaryB}) that statement (\ref{newPcondition}b) is true for a unitary transformation.  To see that
(\ref{newPcondition}a) and (\ref{newPcondition}b) are also sufficient to certify unitarity, note first that
the matrix ${\mathcal B}$ defined in Eq.~(\ref{newPcondition}b) is necessarily Hermitian, because of
the symmetry (\ref{Xisymmetry}) of $\Xi$.  Moreover, 
the normalization condition (\ref{newPcondition}a) implies that the trace of ${\mathcal B}$ is unity.
So the sole non-zero eigenvalue of ${\mathcal B}$ must be 1; that is, ${\mathcal B}$ must be a one-dimensional
projection operator.
Now, if ${\mathcal B}$
is a one-dimensional projection, then there is essentially only a single $\hat{B}$ operator in Eq.~(\ref{transformation}).  (There could be several $\hat{B}_j$'s, but they would all be proportional to each other.) In that case the 
sum condition (\ref{Kraus}), which, as we have seen, follows from the normalization condition (\ref{newPcondition}a), implies that this $\hat{B}$ is unitary.

\section{Continuous Hamiltonian evolution}

In the preceding section, we were interested in a single discrete transformation taking $\hat{\rho}$ to $\hat{\rho}'$.  
We now consider a continuous transformation governed by the von Neumann equation:
\begin{equation}  \label{vonN}
\frac{d\hat{\rho}}{dt} = -\frac{i}{\hbar}[\hat{H},\hat{\rho}],
\end{equation}
where $\hat{H}$ is the Hamiltonian, which we assume to be constant.  
Let $H_\alpha$ be the expansion coefficients of $\hat{H}$ in the phase-point operators $\hat{A}_\alpha$:
\begin{equation}
\hat{H} = \sum_\alpha H_\alpha \hat{A}_\alpha\, ,\hspace{2mm} \hbox{so that}\; H_\alpha = \frac{1}{N} \hbox{Tr}(\hat{H}\hat{A}_\alpha).
\end{equation}
Then we can rewrite Eq.~(\ref{vonN}) in phase space language as
\begin{equation}\label{DWF-dynamics}
    \frac{dW_\alpha}{dt} = \frac{1}{i\hbar} \sum_{\beta\gamma} \Gamma_{\alpha \beta \gamma}\left( H_\beta W_\gamma - W_\beta H_\gamma \right)
\end{equation}
where again $\Gamma_{\alpha\beta\gamma}$ is defined in Eq.~(\ref{Gamma}). Eq.~(\ref{DWF-dynamics}) can be understood as a representation of the discrete Moyal bracket \cite{Klimov, Livine}
\begin{equation}
    \frac{dW_\alpha}{dt} =\frac{1}{i\hbar} \left( H \star W - W \star H \right)_\alpha \equiv \frac{1}{i\hbar} \{\{H,W\}\}_\alpha,
\end{equation}
the star product between the phase-space representation of two operators being defined by $(\hat{B}\hat{C})_\alpha =  \frac{1}{N} \sum_{\beta\gamma} B_\beta C_\gamma \Gamma_{\alpha\beta\gamma} \equiv (B\star C)_\alpha$.

Using the fact that
$\Gamma_{\alpha\gamma\beta} = {\Gamma}^*_{\alpha\beta\gamma}$, we can re-express Eq.~(\ref{DWF-dynamics}) as
\begin{equation}
\frac{dW_\alpha}{dt} = \frac{2}{\hbar} \sum_{\beta\gamma} \hbox{Im}(\Gamma_{\alpha\beta\gamma}) H_\beta W_\gamma  \; .
\end{equation}
Note that this equation can be 
written as
\begin{equation}  \label{Hevolution}
\frac{dW_\alpha}{dt} = \sum_\gamma r_{\alpha\gamma} W_\gamma \; ,
\end{equation}
where
\begin{equation}  \label{rdef}
r_{\alpha\gamma} = \frac{2}{\hbar} \sum_{\beta} \hbox{Im}(\Gamma_{\alpha\beta\gamma}) H_\beta .
\end{equation}
So if we again think of $W_\alpha$ as the probability of the system being at the phase-space point $\alpha$,
then $r_{\alpha\gamma}$ is playing the role of the probability per unit time that a system at the point
$\gamma$ will move to $\alpha$.  We will refer to the $r$'s as transition rates, even though, like the $P$'s of
the preceding section, they can be negative even when $\alpha \ne \gamma$.  (In a classical continuous-time Markov process, $r_{\alpha\gamma}$ can be negative only if $\alpha$ and $\gamma$ are the same, since only in that case 
is $r_{\alpha\gamma}$ not interpreted as a probability per unit time.  We discuss this point further in Section VI.)   In fact, one can see immediately from the definition
(\ref{rdef}) that $r_{\alpha\gamma}$ is antisymmetric in its two indices.  

It is not a coincidence that $r$ is an antisymmetric matrix. As we have seen, the $P$'s describing unitary transformations constitute an orthogonal matrix, and the generators of the orthogonal group are antisymmetric. To see the connection, suppose $P$ describes the transition probabilities corresponding to some differentiable transformation over a short time $\Delta t$ such that 
\begin{equation} \label{infW}
W_\alpha(t + \Delta t) = \sum_\gamma P_{\alpha\gamma} W_\gamma(t).
\end{equation}
Differentiability allows us to expand $P$ to first order in $\Delta t$ as $P_{\alpha\gamma} = \delta_{\alpha\gamma} + s_{\alpha\gamma}\Delta t$ where $s$ is the infinitesimal generator of $P$. The limit $\Delta t \rightarrow 0$ in Eq.~(\ref{infW}) then leads to $dW_\alpha/dt = \sum_\gamma s_{\alpha\gamma} W_\gamma$ which is the same form as Eq.~(\ref{Hevolution}).  So $r$ is the infinitesimal generator of $P$.



Another remarkable property of $r_{\alpha\gamma}$ is that the sum of the transition rates into $\alpha$ from all points in 
phase space is zero:
\begin{equation}
\sum_\gamma r_{\alpha\gamma} = 0,
\end{equation}
as follows directly from Eq.~(\ref{rdef}). More fundamentally, this property is a consequence of Eq.~(\ref{Hevolution}), the normalization of $W$, and the antisymmetry of $r_{\alpha\gamma}$. This does not mean, of course, that the value of $W_\alpha$ does not change---the
rate of change also depends on the values of $W_\gamma$---but it does immediately imply that if $W_\gamma$
is the constant function on phase space (representing the completely mixed state), then it is also constant in time.  
That is, the completely mixed state is unchanged by any Hamiltonian evolution (which is of course correct).  

As in the preceding section, our main concern here is to identify constraints on the transition rates $r_{\alpha\gamma}$ that characterize actual Hamiltonian flows in phase space.  
We begin by inverting Eq.~(\ref{rdef}) so as to express the Hamiltonian function $H_\alpha$ in terms of the $r$'s, if indeed the given set of $r$ values is consistent with a Hamiltonian.

Starting with Eq.~(\ref{rdef}), we use the properties expressed in Eq.~(\ref{properties}) to get
\begin{equation}
\sum_{\alpha\gamma} r_{\alpha\gamma} \hat{A}_\alpha \hat{A}_\gamma = -\frac{i}{\hbar} N^2\left( \sum_\beta
H_\beta \hat{A}_\beta - \frac{1}{N} \sum_\beta H_\beta \right).
\end{equation}
Now multiply by $\hat{A}_\delta$ and take the trace to get
\begin{equation}  \label{firsth}
\frac{i\hbar}{N^2} \sum_{\alpha\gamma} r_{\alpha\gamma} \Gamma_{\alpha\gamma\delta} = H_\delta - \frac{1}{N^2}\sum_\beta H_\beta,
\end{equation}
which gives us the Hamiltonian function $H_\delta$ up to an additive constant.  (The additive constant 
does not affect the dynamics.)  By renaming indices and making use of the symmetries of $\Gamma$ and 
$r_{\alpha\gamma}$, we can re-express Eq.~(\ref{firsth}) somewhat more elegantly as
\begin{equation}  \label{secondh}
H_\beta - \frac{1}{N^2} \sum_\delta H_\delta = \frac{\hbar}{N^2} \sum_{\alpha\gamma} r_{\alpha\gamma} 
\hbox{Im}(\Gamma_{\alpha\beta\gamma}).
\end{equation}

Now, given any candidate set of values $r_{\alpha\gamma}$, Eq.~(\ref{secondh}) will give us some function $H_\beta$ (up to an additive constant).  But not every set of $r$ values actually arises from a Hamiltonian.  To tell whether the given set does represent Hamiltonian evolution, we insert the $H_\beta$ of Eq.~(\ref{secondh}) back into
Eq.~(\ref{rdef}) and see whether that equation yields the same $r$ values we started with.  If so, then
those values do arise from a Hamiltonian; otherwise they do not.  

Carrying out this strategy, we arrive at the following condition characterizing those sets of values $r_{\alpha\gamma}$
that represent Hamiltonian evolution:
\begin{equation}  \label{rcondition}
r_{\alpha\gamma} = \frac{2}{N^2} \sum_{\alpha'\gamma'} \bigg[ \sum_\beta \hbox{Im}(\Gamma_{\alpha\beta\gamma})
\hbox{Im}(\Gamma_{\alpha'\beta\gamma'}) \bigg] r_{\alpha'\gamma'}.
\end{equation}
To write this condition more compactly, let us think of $r_{\alpha\gamma}$ as a column vector with $\alpha\gamma$
as its single index (taking $N^4$ values).  Let us call this column vector $\vec{r}$.  We also define a matrix $R$ in terms of its components as follows:
\begin{equation}  \label{Rdef}
R_{\alpha\gamma,\alpha'\gamma'} = \frac{2}{N^2} \sum_\beta \hbox{Im}(\Gamma_{\alpha\beta\gamma})
\hbox{Im}(\Gamma_{\alpha'\beta\gamma'}).
\end{equation}
Then the condition (\ref{rcondition}) can be re-expressed simply as 
\begin{equation} \label{rconditionshort}
\vec{r} = R\vec{r}.
\end{equation}
In Appendix C we show that the symmetric real matrix $R$ is in fact a projection operator; that is, it has only two distinct eigenvalues, 0 and 1.  
According to Eq.~(\ref{rconditionshort}), a set $\vec{r}$ of transition rates represents a Hamiltonian evolution if and only if it lies in the eigenvalue-1
subspace of $R$.  Moreover, if we start with any real $N^4$-component vector $\vec{v}$ and apply $R$ to 
$\vec{v}$, the result will be a legitimate set of transition rates associated with some Hamiltonian evolution.  (Possibly
the result will be the zero vector, but this vector does indeed define a legitimate set of transition rates.)  

For odd prime values of $N$, the three-point structure function $\Gamma_{\alpha\beta\gamma}$ takes a 
particularly simple form, and we can use this fact to write down the condition (\ref{rcondition}) more explicitly.  
Specifically, we have
\begin{equation}  \label{oddGamma}
\Gamma_{\alpha\beta\gamma} = \frac{1}{N} \exp \left[ -\frac{4 \pi i}{N} \big(\langle\alpha,\beta\rangle + \langle\beta,\gamma\rangle + \langle\gamma, \alpha\rangle\big) \right],
\end{equation}
where again $\langle\alpha, \beta\rangle = \alpha_2 \beta_1 - \alpha_1 \beta_2$.  
Plugging this expression into Eq.~(\ref{rcondition}) and doing the sum over $\beta$, we get
\begin{equation}  \label{rcondition4}
r_{\alpha\gamma} = \frac{1}{N^2} \sum_\zeta (r_{\alpha+\zeta, \gamma + \zeta} - r_{\gamma + \zeta, \alpha+\zeta})
\cos \left[ \frac{4 \pi}{N} \langle\alpha - \gamma, \zeta\rangle \right] .
\end{equation}
If we now allow ourselves to assume that $r_{\alpha\gamma}$ is antisymmetric under interchange of $\alpha$
and $\gamma$, we can combine the two terms in Eq.~(\ref{rcondition4}) to get
\begin{equation} \label{rcondition5}
r_{\alpha\gamma} = \frac{2}{N^2} \sum_\zeta r_{\alpha + \zeta, \gamma + \zeta} \cos \left[ \frac{4 \pi}{N} \langle\alpha - \gamma , \zeta\rangle \right].
\end{equation}
Thus we can take as our condition on the $r$'s either Eq.~(\ref{rcondition4}) by itself, which {\em implies} that
$r_{\gamma\alpha} = - r_{\alpha\gamma}$, or Eq.~(\ref{rcondition5}) {\em together with} the condition
$r_{\gamma\alpha} = - r_{\alpha\gamma}$.  Either of these statements serves to characterize precisely those sets of transition rates
that correspond to Hamiltonian dynamics.  

The case of a single qubit, with $N=2$, is simpler.  In that case, one finds that Eq.~(\ref{rconditionshort}) is equivalent
to a combination of two conditions on the $r$'s that we have already encountered:
\begin{equation} \label{rconditionqubit}
\begin{split}
&\hbox{(a)} \;\; \sum_\alpha r_{\alpha\gamma} = 0. \\
&\hbox{(b)} \;\; r_{\gamma\alpha} = - r_{\alpha\gamma}. 
\end{split}
\end{equation}
In fact one can prove that these conditions are sufficient just by counting the number of free parameters. One finds that of the 16 possible values of an unconstrained $r_{\alpha\gamma}$, only three remain after we impose the conditions in Eq.~(\ref{rconditionqubit}).  This number is the same as the rank of the projection operator $R$ for a qubit: in general, the trace of $R$ is $N^2 - 1$, as we 
show in Appendix C.  Since the linear constraint expressed in Eq.~(\ref{rconditionqubit}) is certainly consistent with the linear constraint in Eq.~(\ref{rconditionshort}), it follows that these constraints are equivalent.  
We note also that for a qubit, three is indeed the number of free parameters in the Hamiltonian, up to an irrelevant additive
constant.

\section{Computing the transition rates}

We now specialize to the case where $N$ is an odd prime.  For such a system, the displacement operators $\hat{D}_\mu$
defined in Eq.~(\ref{displacements}) constitute an orthogonal basis for the space of $N \times N$ matrices \cite{Schwinger}, so 
in particular, we can write the Hamiltonian as a linear combination of them:
\begin{equation}  \label{Hsum}
\hat{H} = \sum_\mu \kappa_\mu \hat{D}_\mu,
\end{equation}
where the $\kappa_\mu$'s are complex numbers.  One can show that $\hat{D}_{(-\mu)} = \hat{D}_\mu^\dag$, so since
$\hat{H}$ is Hermitian, we must have $\kappa_{(-\mu)} = \kappa_\mu^*$.  
It turns out that for each term in the sum (\ref{Hsum}), the corresponding transition rates are
fairly simple, as we are about to see.  Moreover, the $r$'s are linear in $\hat{H}$, so once we have the $r$'s for each term in the
sum, we can add them together to get the transition rates for the whole Hamiltonian.  

In this section, then, we will work out the analogs of transition rates, with the non-Hermitian operator $\hat{D}_\mu$
taking the place of the Hamiltonian.  These can then be combined as in Eq.~(\ref{Hsum}) to get transition rates
for Hamiltonians.    

We begin by finding the phase-space function $D^{(\mu)}_{\beta}$ corresponding to $\hat{D}_\mu$.
That is, we will find
\begin{equation}  \label{d}
D^{(\mu)}_{\beta} = \frac{1}{N} \hbox{Tr} (\hat{D}_\mu \hat{A}_\beta), \hspace{3mm} \hbox{so that} \; \hat{D}_\mu = \sum_\beta D^{(\mu)}_{\beta} \hat{A}_\beta.
\end{equation}
To evaluate $D^{(\mu)}_{\beta}$, we make use of Eq.~(\ref{DtimesD}).
As always, the arithmetic in the exponent of $\omega$ is mod $N$.  Inserting into Eq.~(\ref{d}) 
the definition (\ref{DAs}) of the $\hat{A}$ operators, we have
\begin{equation}
\begin{split}
D^{(\mu)}_{\beta} &= \frac{1}{N} \hbox{Tr} (\hat{D}_\mu \hat{A}_\beta) \\
&= \frac{1}{N^2} \hbox{Tr} \left( \sum_\gamma \hat{D}_\mu \hat{D}_\gamma \omega^{\langle\beta,\gamma\rangle}\right) \\
& = \frac{1}{N^2} \hbox{Tr} \left( \sum_\gamma \hat{D}_{\mu + \gamma} \omega^{\langle\mu,\gamma\rangle/2} \omega^{\langle\beta,\gamma\rangle} \right).
\end{split}
\end{equation}
Now, the trace of $\hat{D}_\alpha$ is zero unless $\alpha$ is zero, so we get a contribution only from the term 
where $\gamma = - \mu$.  This gives us
\begin{equation}  \label{dmu}
D^{(\mu)}_{\beta} = \frac{1}{N} \omega^{\langle\mu,\beta\rangle}.
\end{equation}

We now substitute $D^{(\mu)}_{\beta}$ in place of $H_\beta$ in Eq.~(\ref{rdef}) to get the 
``transition rates'' $r^{(\mu)}_{\alpha\gamma}$ corresponding to the operator $\hat{D}_\mu$:  
\begin{equation}
r^{(\mu)}_{\alpha\gamma} = \frac{2}{\hbar} \sum_{\beta} \hbox{Im}(\Gamma_{\alpha\beta\gamma}) D^{(\mu)}_\beta.
\end{equation}
Using Eq.~(\ref{dmu}) for $D^{(\mu)}_\beta$ and Eq.~(\ref{oddGamma}) for $\Gamma_{\alpha\beta\gamma}$, we get
\begin{equation}
r^{(\mu)}_{\alpha\gamma} =- \frac{2}{\hbar N^2} \sum_{\beta} \sin \left[ \frac{4 \pi}{N} (\langle\alpha,\beta\rangle + \langle\beta,\gamma\rangle + \langle\gamma, \alpha\rangle) \right]\omega^{\langle\mu,\beta\rangle}.
\end{equation}
The sums over $\beta_1$ and $\beta_2$ are straightforward and we find that
\begin{equation}  \label{rmufinal}
r^{(\mu)}_{\alpha\gamma} = \frac{1}{i\hbar} \left[\delta_{\alpha,\gamma+\frac{\mu}{2}} \omega^{2\langle \alpha, \gamma \rangle} - \delta_{\alpha, \gamma-\frac{\mu}{2}} \omega^{-2 \langle \alpha, \gamma\rangle} \right].
\end{equation}
Thus, the contribution to $\hat{H}$ from a specific displacement operator $\hat{D}_\mu$ generates transitions from 
$\gamma$ to $\gamma + \mu/2$ and to $\gamma - \mu/2$.  That is, the displacements effected by the 
transitions are only half as large as the displacement $\mu$.  (But this ``half'' is in the mod $N$ sense.) This factor of one half has been noted before in earlier work where choices of the phase associated with the displacement operator are investigated \cite{Klimov}.

As a simple example, consider a particle that can occupy any of $N$ sites, arranged in a ring, and let the 
Hamiltonian be $\hat{H} = 2 - (\hat{X} + \hat{X}^\dag) = 2 - (\hat{D}_{(1,0)} + \hat{D}_{(-1,0)})$.  This Hamiltonian is analogous to the kinetic energy operator for a 
particle moving on a continuous line.  For example, the eigenstates of $\hat{H}$ are of the form
\begin{equation}  \label{moms}
|p_k\rangle = \frac{1}{\sqrt{N}} \sum_{q=0}^{N-1} \omega^{ k q} |q\rangle, \hspace{1cm} k = 0, \ldots, N-1,
\end{equation}
with eigenvalues $4\sin^2(k \pi/N)$.  When $k \ll N$ these eigenvalues are proportional $k^2$, like the 
eigenvalues of the ordinary kinetic energy operator.  (Note that the constant term, 2, in the Hamiltonian does not
affect the dynamics as expressed in Eq.~(\ref{vonN}) and does not affect the transition rates.)  We take the 
eigenstates of position to be the standard basis, associated with the vertical lines in phase space.  For our choice
of the phase-point operators, this implies that the eigenstates of momentum, 
given in Eq.~(\ref{moms}), are associated with the horizontal lines.  

\begin{figure}[htp]
    \centering
    {\label{fig:5x5-transition-rates}}
  \includegraphics[width=0.45\textwidth]{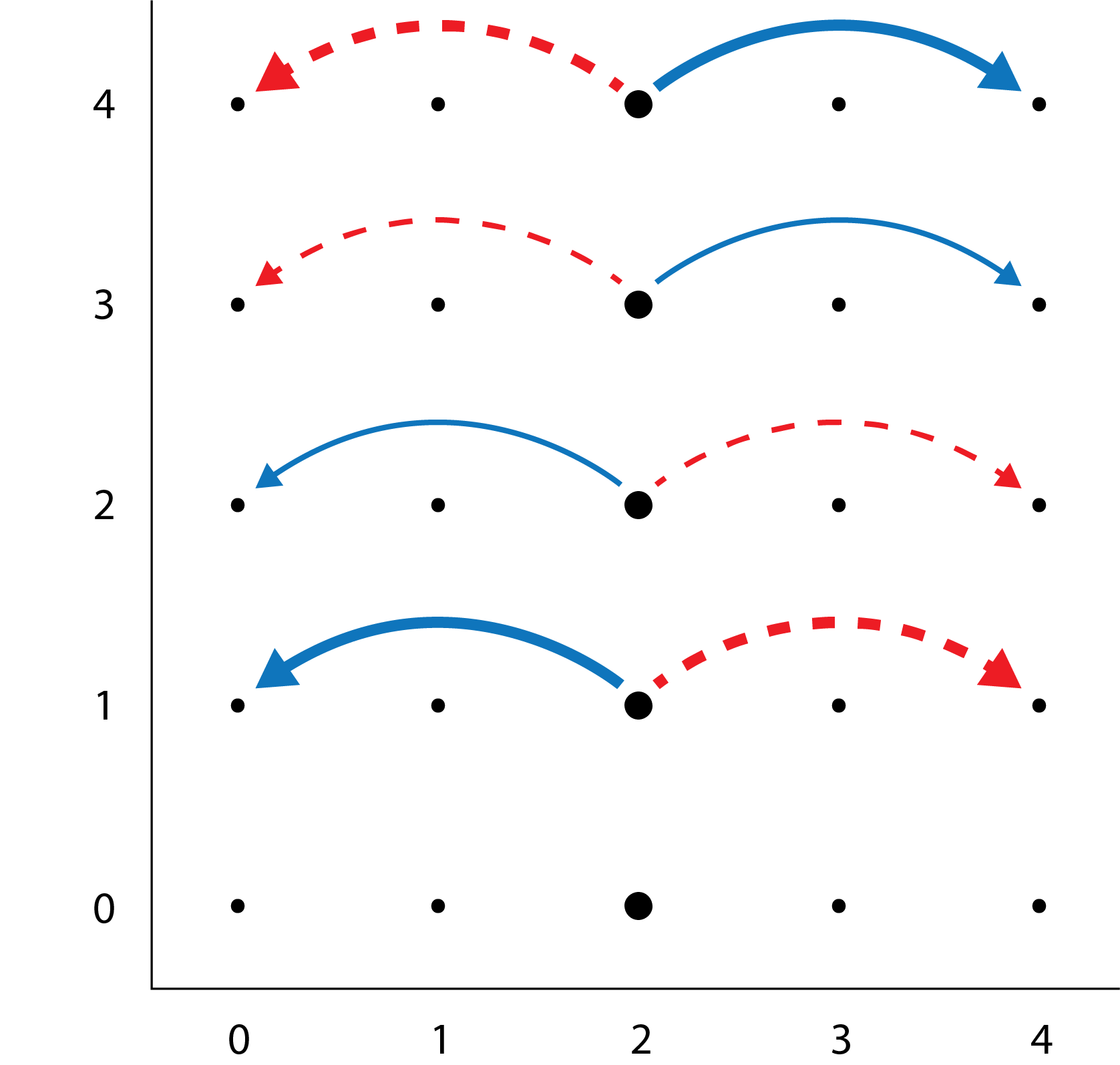}
    \caption{A discrete phase space for a 5-dimensional quantum system. Each of the 25 dots indicates a phase-space point with a discrete Wigner function value of 0 except for the larger dots which have a value of $1/5$; this is the Wigner function for an eigenstate of position with eigenvalue 2. Arrows display transition rates corresponding to the kinetic energy operator for a particle that can occupy 5 sites with periodic boundary conditions. Blue (solid) arrows indicate positive rates and red (dashed) arrows indicate negative rates while the width of an arrow indicates the relative magnitude of the rate. Only the transition rates out of the nonzero Wigner function points are displayed.}
\end{figure}

For this Hamiltonian, according to Eq.~(\ref{rmufinal}) the transition rates $r^H$ are
\begin{equation}  \label{exampler}
\begin{split}
r^H_{\alpha\gamma} &=  -r^{(1,0)}_{\alpha\gamma} - r^{(-1,0)}_{\alpha\gamma} \\
&= -\frac{2}{\hbar} (\delta_{\alpha, \gamma + \eta} + \delta_{\alpha, \gamma - \eta} ) \sin\left(\frac{4 \pi}{N} \langle \alpha, \gamma \rangle \right) \\
&= \frac{2}{\hbar}  (\delta_{\alpha, \gamma + \eta}  -  \delta_{\alpha, \gamma - \eta} ) \sin\left( \frac{2 \pi}{N} \gamma_2 \right),
\end{split}
\end{equation}
where $\eta = ((N+1)/2, 0)$.  So if a particle could start at a specific phase-space point $\gamma$, it would, to first
order in time, move to the two points farthest from $\gamma_1$ on the circle and not change its momentum coordinate
$\gamma_2$ at all.  Of course a system cannot start in such a state.  If it starts in an eigenstate of position---for definiteness let us say it starts at $\gamma_1 = 2$---then its initial Wigner function is uniform over the vertical line $\gamma_1 = 2$. An example of such a scenario for a ring with $N = 5$ sites is shown in Fig. 1. To first order in time, the contribution from each point on this line moves halfway around the circle, to the points $2+(N+1)/2$ and 
$2+(N-1)/2$, but because of the factor $\sin(2 \pi \gamma_2/N)$ in Eq.~(\ref{exampler}), when we sum over $\gamma_2$ to get the probability distribution over position, we find that it has not changed at all.  And indeed, starting from a 
position eigenstate, the distribution of positions should not change at all to first order in time.  (The change is of second
order.)

\section{Conclusions}

An ordinary stochastic process on an $N \times N$ grid of points would be defined by specifying, for each
pair of points $(\alpha, \gamma)$, the
probability $P_{\alpha\gamma}$ that the system will make the transition to the point $\alpha$ if it is currently
at the point $\gamma$.  The only constraints on these probabilities are 
\begin{equation}  \label{classicalconstraints}
\begin{split}
&\hbox{(a)}\;\; \sum_\alpha P_{\alpha\gamma} = 1 \hspace{3mm} \hbox{for each $\gamma$}; \\
&\hbox{(b)} \;\;P_{\alpha\gamma} \ge 0 \hspace{3mm} \hbox{for each pair $(\alpha, \gamma)$}. 
\end{split}
\end{equation}
We have seen that for a quantum process described in discrete phase space, the constraints are different.
We still have the normalization constraint of Eq.~(\ref{classicalconstraints}a), but Eq.~(\ref{classicalconstraints}b) is replaced 
by a different positivity condition, namely, that the matrix
\begin{equation}  \label{newconditionb}
{\mathcal B}_{\beta\delta} = \frac{1}{N^2} \sum_{\alpha\gamma} \Xi_{\beta\alpha\delta\gamma} P_{\alpha\gamma} ,
\end{equation}
in which $\beta$ and $\delta$ are understood to be the matrix indices, is positive semidefinite.  Here
$\Xi_{\beta\alpha\delta\gamma}$ is a complex-valued function of its four arguments, but as we see
in Appendix A, it is a fairly simple function when $N$ is an odd prime.  It is nonzero only when its arguments
form a parallelogram in the discrete phase space, and in that case its magnitude is always unity and its phase is proportional to the parallelogram's area.  Note that both the classical stochastic process and the general quantum 
process allow the same number of free parameters, namely, $N^2(N^2 - 1)$.  It is only the inequalities constraining these parameters that are different.  

For the special case in which the quantum process is a unitary transformation, the condition that ${\mathcal B}$ be positive-semidefinite
can be replaced by the stronger requirement that ${\mathcal B}$ be of rank one (in which case
the sole non-zero eigenvalue must be 1 in order for the normalization condition (\ref{classicalconstraints}a) to be satisfied).  It is interesting to count parameters in this case as well.  If we ignore normalization for now, it takes
$2N^2-1$ real numbers to specify an $N^2 \times N^2$ rank-one Hermitian matrix ${\mathcal B}_{\beta\delta}$.  (It takes $N^2$ complex numbers, or $2N^2$ real numbers, to specify a vector $B_\beta$ from which ${\mathcal B}_{\beta\delta}$ is constructed
via ${\mathcal B}_{\beta\delta} = B_\beta B_\delta^*$, but one of those real numbers is the overall phase of $B$, which is 
lost in ${\mathcal B}$.)  Imposing the $N^2$ normalization equations in Eq.~(\ref{classicalconstraints}a) then leaves us
with $N^2 - 1$ real parameters, which is indeed the number of parameters required to specify a special unitary 
transformation in an $N$-dimensional Hilbert space.  (An overall phase of the unitary transformation does not 
affect the evolution of the density matrix and therefore does not affect our transition probabilities.)  

We now turn to the case of a continuous transformation.  An ordinary continuous-time Markov process can be
described by a set of differential equations of the form
\begin{equation}  \label{classicalmaster}
\frac{dW_\alpha}{dt} = \sum_{\gamma} r_{\alpha\gamma} W_\gamma,
\end{equation}
where $W_\gamma$ is the probability that the system is in the state $\gamma$, and the transition rate $r_{\alpha\gamma}$, for $\alpha \ne \gamma$, is the probability per unit time that a system in the state
$\gamma$ will make a transition to $\alpha$.  The quantity $r_{\alpha\alpha}$ is the negative of the probability
per unit time that a system in the state $\alpha$ will {\em leave} that state.  Any set of transition rates is allowed that satisfy the following
two constraints.
\begin{equation} \label{classicalMarkov}
\begin{split}
&\hbox{(a)} \;\; \sum_\alpha r_{\alpha\gamma} = 0. \\
&\hbox{(b)} \;\; \hbox{For $\alpha \ne \gamma$,} \; r_{\alpha\gamma} \ge 0.
\end{split}
\end{equation}
The first of these conditions follows directly from the requirement that the probability distribution $W_\alpha$ 
remain normalized no matter what that distribution might be.  The second requirement follows from the assumption
that any probability must be non-negative.  

In the quantum case, for Hamiltonian evolution, the discrete Wigner function $W_\alpha$ follows a set of 
differential equations of the same form as in Eq.~(\ref{classicalmaster}), but the constraints are different.  
Not surprisingly, these constraints allow fewer free
parameters than Eq.~(\ref{classicalMarkov}), just as the 
unitary conditions considered above allow fewer parameters than the classical rules (\ref{classicalconstraints}) or the rules for a general trace-preserving quantum transformation.   
We have seen that for Hamiltonian evolution, a vector of transition rates $\vec{r}$ is allowed if and only if
$R \vec{r} = \vec{r}$, where the projection operator $R$ is defined in Eq.~(\ref{Rdef}). 
This requirement implies two others:
\begin{equation}  \label{quantumcontinuousconclusions}
\begin{split}
&\hbox{(a)} \;\; \sum_\alpha r_{\alpha\gamma} = 0. \\
&\hbox{(b)} \;\; r_{\gamma\alpha} = - r_{\alpha\gamma}.
\end{split}
\end{equation}
The first of these is the familiar normalization-preserving constraint.  The second is completely foreign to
the classical picture.  First, it forces any non-trivial evolution to violate Eq.~(\ref{classicalMarkov}b).  It also 
forces $r_{\alpha\alpha}$ to be zero.  This latter fact {\em would} mean that a system in state $\alpha$ could not 
leave that state, if it were not for the fact that some values of $r_{\alpha\gamma}$ are negative.
A negative transition rate from $\gamma$ to $\alpha$ reduces the value of $W_\alpha$, but
at a rate proportional to $W_\gamma$, not to $W_\alpha$.  

For a single qubit, the two conditions in Eq.~(\ref{quantumcontinuousconclusions}) are equivalent to $R \vec{r} = \vec{r}$ and are therefore sufficient to determine what sets of transition rates are allowed.  For the case where $N$ is an 
odd prime, we need an additional condition:
\begin{equation}
r_{\alpha\gamma} = \frac{2}{N^2} \sum_\zeta r_{\alpha + \zeta, \gamma + \zeta} \cos \left[ \frac{4 \pi}{N} \langle\alpha - \gamma , \zeta\rangle \right].
\end{equation}
This equation, like the form (\ref{XiAppendixA}) of the four-point structure function $\Xi$ or the form (\ref{rmufinal})
of the ``transition rates'' associated with a displacement operator, highlights the 
important role of the 
symplectic product for the odd-prime case.  

It is worth commenting further on the significance of the symplectic product.  It is well known that when $N$ is 
an odd prime,
any unit-determinant linear transformation acting on the phase space, regarded as a two-dimensional vector space over the $N$-element field, is equivalent 
to a unitary transformation acting on the phase-point operators (see, for example, Refs.~\cite{Vourdas2, Appleby}).  That is, if $L$ is a 
unit-determinant linear transformation, then there is a corresponding unitary $\hat{U}_L$ such that 
$\hat{A}_{L\alpha} = \hat{U}_L \hat{A}_\alpha \hat{U}_L^\dag$ for all points $\alpha$.  This means that
the basic structure of the theory is unchanged by such a transformation. (For example, the forms of Eqs.~(\ref{pureW}) and (\ref{DWF-dynamics}) are unchanged.)  These special linear 
transformations---symplectic transformations---do
not preserve any non-trivial distance function or any notion of angle, but they do preserve the symplectic 
product, which can be taken to define a notion of area, as we discuss in Appendix A.  This fact is roughly analogous to 
the fact that in classical mechanics, phase-space volume is preserved under canonical transformations.  

The appearance of negative probabilities, both in the discrete Wigner function itself and in the transition probabilities
and transition rates, would be more disturbing if it were not for the fact that these non-standard probabilities
are always associated with {\em illegal} states.  For example, we speak of a negative transition rate from some
phase-space point $\gamma$ to another phase-space point $\alpha$.  But in standard quantum theory, the 
system cannot actually {\em be} at the point $\gamma$ and cannot go to the point $\alpha$.  The rules we have derived that limit the 
sets of allowed transition rates and transition probabilities, together with the rules restricting the Wigner function, evidently entail restrictions that force the probabilities 
of all {\em observable} events to be non-negative.
An interesting question for future research is whether
the constraints we have noted here can all be derived,
within a minimal framework, simply by requiring non-negativity 
at this level.

\section*{Acknowledgements}

W.F.B. gratefully acknowledges valuable discussions with Miles Blencowe, Peter Johnson, and Apostolos Vourdas.

\section*{Appendix A:  The four-point structure function $\Xi_{\alpha\beta\gamma\delta}$}

Here we evaluate the four-point structure function 
\begin{equation}
\Xi_{\alpha\beta\gamma\delta} = \frac{1}{N} \hbox{Tr} (\hat{A}_\alpha \hat{A}_\beta \hat{A}_\gamma \hat{A}_\delta)
\end{equation}
for the case when $N$ is an odd prime.  
From the definition
\begin{equation}
\hat{A}_\alpha = \frac{1}{N} \sum_\mu \hat{D}_\mu \, \omega^{\langle \alpha, \mu \rangle },
\end{equation}
we have
\begin{equation}  \label{bigXi}
\Xi_{\alpha\beta\gamma\delta} = \frac{1}{N^5} \sum_{\mu\nu\rho\sigma} \hbox{Tr} ( \hat{D}_\mu \hat{D}_\nu \hat{D}_\rho \hat{D}_\sigma)
\omega^{(\langle \alpha,\mu \rangle + \langle \beta, \nu \rangle + \langle \gamma, \rho\rangle + \langle \delta, \sigma\rangle)}.
\end{equation}
Now we use the composition rule (\ref{DtimesD}) for displacement operators to get
\begin{equation}
\hat{D}_\mu \hat{D}_\nu \hat{D}_\rho \hat{D}_\sigma = \hat{D}_{\mu + \nu + \rho + \sigma} \omega^{(\langle \mu, \nu + \rho + \sigma\rangle + \langle \nu, \rho+\sigma\rangle + \langle \rho, \sigma \rangle )/2}.
\end{equation}
The trace of $\hat{D}_{\mu + \nu + \rho + \sigma}$ is $N\delta_{\mu + \nu + \rho + \sigma, 0}$, so one of the sums
in Eq.~(\ref{bigXi}) can be done immediately.  In the remaining sums, we use, a few times, the fact that
\begin{equation}
\sum_{x=0}^{N-1} \omega^{xy} = N\delta_{y,0}.
\end{equation}
The final result can be written as
\begin{equation}  \label{XiAppendixA}
\Xi_{\alpha\beta\gamma\delta} = \delta_{\alpha-\delta, \beta- \gamma} \omega^{2\langle \delta-\alpha, \beta-\alpha\rangle}.
\end{equation}
The Kronecker delta forces the points $\alpha$, $\beta$, $\gamma$, $\delta$ to be the corners of a parallelogram---possibly a degenerate parallelogram in which all the vertices lie on a single line---and 
the exponent of $\omega$ can be interpreted as twice the area of the parallelogram.  (If we picture the phase space as a lattice with unit spacing between neighboring points, this area is equal to the ordinary signed area in the plane, evaluated mod $N$.  The sign is positive if the path $\alpha \rightarrow \beta \rightarrow \gamma \rightarrow \delta \rightarrow \alpha$ is counter-clockwise.)  Thus $\Xi_{\alpha\beta\gamma\delta}$ is zero for most values of 
its indices.  For any given values of $\alpha$, $\beta$, and $\gamma$, there is only one value of $\delta$ for 
which $\Xi_{\alpha\beta\gamma\delta}$ is not zero.

\section*{Appendix B: Inverting the formula for $P_{\alpha\gamma}$}

Recall that the transition probabilities $P_{\alpha\gamma}$ are given in terms of ${\mathcal B}_{\beta\delta}$
by Eq.~(\ref{Pdef}):
\begin{equation} \label{Pdefagain}
P_{\alpha\gamma} = \frac{1}{N} \sum_{\beta\delta} \hbox{Tr}(\hat{A}_\alpha \hat{A}_\beta \hat{A}_\gamma \hat{A}_\delta) {\mathcal B}_{\beta\delta}.
\end{equation}
Here we wish to invert this equation to get an expression for ${\mathcal B}_{\beta\delta}$.
We begin by recalling Eq.~(\ref{properties}): 
for any $N \times N$ matrix $\hat{M}$,
\begin{equation} \label{propertiesagain}
\begin{split}
&\hbox{(a)} \;\; \sum_\alpha \hat{A}_\alpha \hbox{Tr}(\hat{M}\hat{A}_\alpha) = N\hat{M} \\
&\hbox{(b)} \;\; \sum_\alpha \hat{A}_\alpha \hat{M} \hat{A}_\alpha = N(\hbox{Tr}\hat{M})\hat{I}.
\end{split}
\end{equation}
Starting with Eq.~(\ref{Pdefagain}), we multiply both sides by $\hat{A}_\gamma$ and use Eq.~(\ref{propertiesagain}a) to 
get
\begin{equation}
\sum_\gamma P_{\alpha\gamma}\hat{A}_\gamma = \sum_{\beta\delta} {\mathcal B}_{\beta\delta} \hat{A}_\delta \hat{A}_\alpha \hat{A}_\beta.  
\end{equation}
Now multiply on the left by $\hat{A}_\nu$ and on the right by $\hat{A}_\mu \hat{A}_\alpha$, sum over $\alpha$ and use 
Eq.~(\ref{propertiesagain}b):
\begin{equation}
\sum_{\alpha\gamma} P_{\alpha\gamma} \hat{A}_\nu \hat{A}_\gamma \hat{A}_\mu \hat{A}_\alpha = N \sum_{\beta\delta} {\mathcal B}_{\beta\delta} \hat{A}_\nu \hat{A}_\delta \hbox{Tr}(\hat{A}_\beta \hat{A}_\mu).
\end{equation}
Finally, take the trace of both sides and use the fact that $\hbox{Tr}(\hat{A}_\nu \hat{A}_\delta) = N\delta_{\nu\delta}$ to get
\begin{equation}  \label{inverseagain}
{\mathcal B}_{\mu\nu} = \frac{1}{N^3} \sum_{\alpha\gamma} \hbox{Tr}(\hat{A}_\mu \hat{A}_\alpha \hat{A}_\nu \hat{A}_\gamma) P_{\alpha\gamma} = \frac{1}{N^2} \sum_{\alpha\gamma} \Xi_{\mu\alpha\nu\gamma} P_{\alpha\gamma}.
\end{equation}
This is the desired equation for ${\mathcal B}_{\beta\delta}$.

\section*{Appendix C: Showing that $R$ is a projection}

Here we want to show that the matrix $R$ defined by
\begin{equation}  \label{whatRis}
R_{\alpha\gamma, \alpha'\gamma'} = \frac{2}{N^2} \sum_\beta \hbox{Im}(\Gamma_{\alpha\beta\gamma})
\hbox{Im}(\Gamma_{\alpha' \beta \gamma'})
\end{equation}
is a projection operator.  Again,
we are thinking of the pair $\alpha\gamma$ as a single matrix index taking $N^4$ values.  The matrix is clearly real and symmetric, so we need only show that $R^2 = R$.  

We begin by noting the following fact about $\hbox{Im}(\Gamma)$.
\begin{equation}  \label{lemma}
\frac{2}{N^2} \sum_{\alpha\gamma} \hbox{Im}(\Gamma_{\alpha\beta\gamma}) \hbox{Im}(\Gamma_{\alpha\beta' \gamma} ) = - \frac{1}{N^2} + \delta_{\beta\beta'}.
\end{equation}
One can see that this is true by writing out $\hbox{Im}(\Gamma)$ in terms of traces of products of $A$ matrices
and then using the two properties given in Eq.~(\ref{properties}).  

We want to show that
\begin{equation}
\sum_{\alpha'\gamma'} R_{\alpha\gamma, \alpha' \gamma'} R_{\alpha' \gamma', \alpha'' \gamma''} = R_{\alpha \gamma, \alpha'' \gamma''}.
\end{equation}
Using the definition (\ref{whatRis}) and letting $G_{\alpha\beta\gamma} = \hbox{Im}(\Gamma_{\alpha\beta\gamma})$, we can write the left-hand side as
\begin{equation}
\frac{4}{N^4} \sum_{\alpha' \gamma'} \bigg( \sum_\beta G_{\alpha\beta\gamma}
G_{\alpha'\beta\gamma'} \sum_{\beta'} G_{\alpha' \beta' \gamma'} G_{\alpha'' \beta' \gamma''} \bigg). 
\end{equation}
Now doing the sum over $\alpha'$ and $\gamma'$ and invoking Eq.~(\ref{lemma}), we can rewrite this as
\begin{equation}
\frac{2}{N^2} \sum_{\beta \beta'} G_{\alpha\beta\gamma} \bigg( -\frac{1}{N^2} + \delta_{\beta\beta'} \bigg)
G_{\alpha'' \beta' \gamma''}.
\end{equation}
The term with $1/N^2$ yields zero, because the imaginary part of $\Gamma$ vanishes when we sum over 
one of the indices.  So we are left with 
$
\frac{2}{N^2} \sum_\beta G_{\alpha\beta\gamma} G_{\alpha'' \beta \gamma''}
$,
which equals $R_{\alpha\gamma, \alpha'' \gamma''}$.  This is what we wanted to show.  

\medskip

Finally, we will find it useful to know the dimension of the subspace onto which $R$ projects.  This is given
by the trace of $R$, that is, 
$\sum_{\alpha\gamma} R_{\alpha\gamma, \alpha\gamma}$, which we
can write as
\begin{equation} 
-\frac{1}{2N^4} \sum_{\alpha\beta\gamma} \left[ \hbox{Tr}(\hat{A}_\alpha \hat{A}_\beta \hat{A}_\gamma) - \hbox{Tr}(\hat{A}_\alpha \hat{A}_\gamma \hat{A}_\beta)\right]^2.
\end{equation}
Using the properties given in Eq~(\ref{properties}), we find that

\begin{equation}
\begin{split}
&\sum_{\alpha\beta\gamma} \hbox{Tr}(\hat{A}_\alpha \hat{A}_\beta \hat{A}_\gamma) \hbox{Tr} (\hat{A}_\alpha \hat{A}_\beta \hat{A}_\gamma) = N^4, \;\; \hbox{and} \\
&\sum_{\alpha\beta\gamma} \hbox{Tr}(\hat{A}_\alpha \hat{A}_\beta \hat{A}_\gamma) \hbox{Tr} (\hat{A}_\alpha \hat{A}_\gamma \hat{A}_\beta) = N^6.
\end{split}
\end{equation}
It follows that $\hbox{Tr} \, R = N^2 - 1$.  So $R$ projects onto a subspace of dimension $N^2 - 1$ (as it should,
since this is the number of parameters needed to specify a Hamiltonian, up to an additive constant).

\end{document}